\documentclass{article}
\usepackage{amssymb}

\usepackage{amsfonts}
\usepackage{sw20ssur}

\newtheorem{theorem}{Theorem}
\newtheorem{acknowledgement}[theorem]{Acknowledgement}

\newtheorem{conjecture}[theorem]{Conjecture}

\input{tcilatex}

\begin{document}

\title{Constraints On Cosmic Dynamics}
\author{Manasse R. Mbonye \\
\\
\textit{Department of Physics, }\\
\textit{Rochester Institute of Technology, }\\
\textit{64 Lomb Drive, Rochester, NY 14623.} }
\date{The Date}
\maketitle

\begin{abstract}
Observationally, the universe appears virtually critical. Yet, there is no
simple explanation for this state. In this article we advance and explore
the premise that the dynamics of the universe always seeks equilibrium
conditions. Vacuum-induced cosmic accelerations lead to creation of
matter-energy modes at the expense of vacuum energy. Because they gravitate,
such modes constitute inertia against cosmic acceleration. On the other
extreme, the would-be ultimate phase of local gravitational collapse is
checked by a phase transition in the collapsing matter fields leading to a
de Sitter-like fluid deep inside the black hole horizon, and at the expense
of the collapsing matter fields. As a result, the universe succumbs to
neither vacuum-induced run-away accelerations nor to gravitationally induced
spacetime curvature singularities. Cosmic dynamics is self-regulating. We
discuss the physical basis for these constraints and the implications,
pointing out how the framework relates and helps resolve standing puzzles
such as "why did cosmic inflation end?", "why is $\Lambda $ small now?" and
"why does the universe appear persistently critical?". The approach does, on
the one hand, suggest a future course for cosmic dynamics, while on the
other hand it provides some insight into the physics inside black hole
horizons. The interplay between the background vacuum and matter fields
suggests an underlying symmetry that links spacetime acceleration with
spacetime collapse and global (cosmic) dynamics with local (black hole)
dynamics.
\end{abstract}

\section{\protect\bigskip Introduction}

Cosmology is moving into a new chapter following the observational evidence
[1] that the universe is currently accelerating. Cosmic dynamics can no
longer be considered to result simply from a post-inflationary free
expansion of spacetime, modulated by gravity. Notwithstanding, one still
expects that the events leading to the current dynamical state of the
universe are rooted in its early past and more and more understanding of the
past will certainly help sort out the present puzzles.

One such standing puzzle can be framed in the question "why did the early
cosmic inflation end?". Over the years following the introduction of the
inflationary model of the early universe [2], various potentials have been
constructed that seek to "gracefully" take the universe out of this initial
de Sitter phase. To a varying extent, the constructions do achieve their aim
of reconciling the events of the early universe with currently observed
phenomena such as large scale structure and CMB anisotropy. Such
constructions are, however, essentially phenomenological in nature, tailored
to conform with a unique and known result. Of course this does not diminish
the vital role the techniques have played and the consequent successes. In
theory, though, the universe could have continued to inflate eternally. The
physics of "why did the early cosmic inflation end?" is still not clear.
Cosmology will have to address this question, in part, to provide a
foundational framework for dealing with the current and any future cosmic
acceleration scenarios. Interestingly, this question may be non-trivially
related (see below) to the old Cosmological Constant Problem and also to
some newly emerging puzzles. To date, for example, it is still not
understood whether or not dark energy, the driver of the current cosmic
acceleration, is in any way related to the field(s) responsible for cosmic
inflation of the early universe. If the two are related (which is the
position taken in this work) then a resolution of the `old' Cosmological
Constant Problem may also help resolve the current Coincidence Problem (and
vise versa). Beyond this, understanding the characteristics of dark energy
is a necessary prerequisite to plotting the future dynamics of the universe.
The relevant equation of state, (and indeed the general evolutionary
character of the universe) is dependent, among other things, on whether or
not dark energy does interact with matter fields and, if so, what the nature
and extent of the interaction(s) is/are. In the end one wants to answer the
following question: will the current universe lock itself into (the first
ever) run-away\footnote{%
A runaway situation exists if, for example, $\Lambda $ is constant (as some
models suggest) since the density of the matter fields always redshifts with
cosmic expansion} acceleration, in future, or will it find an exit as it did
before (during the inflationary era)? The present lack of an answer to these
questions constitutes what has been referred to as [3] \textit{a dynamical
problem}.

In a recent article [3] it was pointed out how the current cosmic
acceleration does find an exit in future and how this behavior is related to
a previous exit out of the inflationary era of the early universe. In both
cases, the central feature is that cosmic (spacetime) acceleration always
triggers the creation, from the background vacuum energy, of matter-energy
(modes) fields which increase the universes' gravitational content. The
behavior leads to an increase in spacetime inertia against vacuum-induced
cosmic acceleration. This, in our approach, forms the physical basis for
cosmic exit out of inflation (and indeed, out of any consequent
accelerations). The justification is based on the following Cosmic
Equilibrium Conjecture (CEC) we recently put forward which states that:

\begin{conjecture}
The universe is always in search of dynamical equilibrium\footnote{%
Dynamical equilibrium here (and henceforth) refers to the dynamical state of
a critical universe}, and will back-react to vacuum-induced tendencies to
accelerate it, by increasing its inertia (at the expense of the vacuum
energy).
\end{conjecture}

The conjecture implies a universe with certain characteristics which make
cosmic dynamics relatively predictable. The aim of this article is, first to
discuss the physical basis for (i.e. the character of the universe implied
by) CEC, consistent with the observed universe. Then we point out how CEC
resolves some current puzzles and ask what predictions it makes. Thereafter
we generalize the conjecture into one that applies to any (and all)
tendencies to shift the universe from equilibrium. The motivation for this
generalization is straightforward. Besides spacetime stretching under the
influence of vacuum energy, there are other ways cosmic equilibrium could be
compromised. For example, spacetime could undergo total collapse under
gravitational influences. We study the consequences of the generalized
conjecture to find that it (i) is consistent with the observed universe;
(ii) explains some of the current puzzles and (iii) is predictive.

The rest of the paper is organized as follows. In Section 2 we put forward a
case for cosmic stability against vacuum-induced run-away accelerations and
study the consequences. In Section 3 we discuss cosmic stability against
gravitationally-induced run-away spacetime collapse. We then make a general
case for a universe that self-regulates its dynamics. In Section 4 the paper
is concluded

\section{Case against run-away cosmic accelerations}

\subsection{Background}

We begin by justifying the physical basis for CEC through the character of
the universe it implies. With regard to spacetime acceleration, one feature
that lends support to the influence of CEC has to do with the
characteristics of the early de Sitter-like phase of the universe, and the
fate of the cosmological `constant' parameter $\Lambda $. Traditional
field-theoretic considerations, based on the characteristics ascribed to de
Sitter vacua, have generally required that the large potential energy $\sim
10^{94}gm~cm^{-3}$ of the vacuum responsible for the early cosmic inflation
should still be around, in the same form. The virtual absence of this energy
leads to the traditional Cosmological Constant Problem. Here, we give a
brief comment on this issue. Notwithstanding its rich symmetry, which gives
it a mathematical beauty, physically a de Sitter spacetime can not be
time-invariant. To appreciate this one need only recall that such a geometry
is known to possess a spacetime horizon at $\Lambda ^{-\frac{1}{2}}$.
Associated with such a horizon is a temperature $T_{\Lambda }=\frac{1}{2\pi }%
\left( \frac{\Lambda }{3}\right) ^{\frac{1}{2}}$ which should give rise to a
matter-energy radiation flux [4] (of power $\frac{dE}{dt}\ $from the area$\
A_{\Lambda }$) i.e. $A_{\Lambda }\frac{dE}{dt}\varpropto T_{\Lambda
}^{4}\varpropto \Lambda ^{2}$. The issue of whether such a radiation is
physical, and the potential implications, has for sometime now been quite
unclear. Lately researchers are revisiting this problem. For example, Parikh
[5] has demonstrated that Hawking radiation from a de Sitter horizon can be
put on the same footing as that from a black hole, in the sense that both
result from a semi-classical tunneling process from the respective horizons.
This view has recently been generalized by Medved [6] to apply to arbitrary
dimensions. The potential implication that the idealized `eternal' de Sitter
state may not be entirely physical has also been raised lately. For example,
in a recent article "Trouble with de Sitter space", N. Goheer, M. Kleban and
L. Susskind [7] have argued on general grounds that the de Sitter space can
not persist for times of order of the Poincare recurrence time. Citing
previous works of other authors [8] to support their claim, they conclude
that it is possible an eternal de Sitter space does not exist, because there
are always instabilities which cause the space to decay in a time shorter
than the recurrence time.

Such arguments\ lend credence to our assertion above that a physical de
Sitter spacetime can not be time-invariant. Rather, soon as it comes into
existence, the de Sitter state evolves a horizon-induced time-asymmetric
behavior. Such behavior has been previously suggested by Padmanabhan [9]. In
this respect therefore, it follows that the cosmological `constant' $\Lambda 
$ of the early universe (responsible for the inflationary cosmic
acceleration) must decay with time into a radiation flux of matter fields%
\footnote{%
The argument implies that there may, in fact, be no cosmological constant
problem.}. In our model, the created matter fields increase the inertia of
the universe by increasing the latter's gravitational content, and at the
expense of the energy in the cosmological constant. Such particle flux $\sim
\Lambda ^{2}$ can be quite large in the early universe when $\Lambda $ is
known to be large.

A parallel, and possibly more illustrative, approach is to treat the
universe as a cosmic engine. In this picture, a dominating vacuum energy $%
\frac{\Lambda }{8\pi G}$ fuels the universe, doing mechanical work on the
spacetime by accelerating it. However, like any other engine, the universe
can not convert all its fuel to pure work. Some of the fuel must be
dissipated to create/increase \textit{cosmic} inertia or internal energy.
One identifies this cosmic internal energy as modes of radiation (matter)
fields. Such dissipation of the (implied interacting) vacuum energy is
clearly similar to the above mentioned time-asymmetric evolution of the
(physical) de Sitter phase.

There are several consequences to this scenario, all of which justify the
(initial) physical basis for CEC. First as already mentioned, the created
matter fields become a source of cosmic inertia through their gravitating
nature. Such matter creation, coupled with the consequent decay of $\Lambda $%
, compromises the existing vacuum-induced cosmic acceleration, driving the
universe out of the inflationary and subsequent accelerations. In our model
this provides the physically justified answer to "why did the early
inflation end?". Through its time-asymmetric character, the de Sitter state
dictates its own end. The universe is thereafter propelled into a radiation
dominated state. Finally, the decay of $\Lambda $ offers a venue for
discussing "why is $\Lambda $ small now?"\footnote{%
In a different approach, Brandenberger has argued [10] that gravitational
back-reaction may lead to a dynamical cancellation mechanism for a bare
cosmological constant by giving rise to an evolving anti-de Sitter-like
field.} and, by extension, why the vacuum energy density $\rho _{v}$ is
coincidentally of the same order as that of the matter fields.

The framework implies that the post-inflationary radiation/matter dominated
state can not be permanent, either. After inflation, when the universe
becomes radiation/matter dominated it continues to expand, mostly under its
own momentum. The density of the matter fields $\rho _{m}\left( a\left(
t\right) \right) $ redshifts with the scale factor $a\left( t\right) $
(modulated only by gravity) and follows the standard dilution power law of $%
-3\gamma =\{-4,-3\}$, depending on whether the fields are relativistic or
cold and pressureless. Eventually $\rho _{m}$ falls below $\rho _{v}$ and
the vacuum energy density dominates the universe again, albeit at a much
lower value, leading to a weaker cosmic acceleration as observed currently.
Applying the preceding reasoning once again, one can infer (as we find
below) that, in future, the universe must exit out of the current low
acceleration because of the creation of long wavelength modes. Although such
modes $\sim \frac{1}{\sqrt{\Lambda }}$ may be too weak to detect currently
they, nevertheless, gravitate. It also follows that the current cosmic
acceleration must be variable.

In this model, such an interplay between the background vacuum energy and
matter fields does underlie cosmic dynamics. The approach presents a history
of cosmic dynamics consistent with both standard theoretical predictions
e.g. inflation, and with observations e.g. the current cosmic acceleration,
and the critical state of the universe. At the same time it suggests
solutions to long standing puzzles, as mentioned above. Moreover, the
approach offers a resolution to the \textit{dynamical problem} by making the
future dynamics of the universe predictable. A quantitative discussion of
these issues can be found in [3]. Presently, we only wish to point out the
implications of CEC and in this regard, we will highlight some of the ideas
in [3] leading to the solution for the evolution of $\rho _{v}$ and briefly
point out those features of $\rho _{v}$ relevant to the current discussion.
Later, we generalize the conjecture to deal with all tendencies to shift
cosmic dynamics from equilibrium and discuss the implications.

\subsection{Framework}

Generating the cosmic background vacuum (or dark) energy $\frac{\Lambda }{%
8\pi G}$ is a cosmological parameter $\Lambda $ which in this model takes on
the functional form 
\begin{equation}
\Lambda \left( t\right) =m_{pl}^{4}\left( \frac{a_{pl}}{a(t)}\right)
^{\sigma \left( t\right) }e^{-\tau H}=\Lambda _{pl}\left( \frac{a_{pl}}{a(t)}%
\right) ^{\sigma \left( t\right) }e^{-\tau H},  \tag{2.1}
\end{equation}%
where $m_{pl}$ is the Planck mass, $a_{pl}$ (the fluctuation scale) is the
size scale of a causally connected region of space at the Planck time $t_{pl%
\text{, }}$and $\tau $ is of order of the Planck time. Further, $H$ is the
Hubble parameter and $a\left( t\right) $ is the cosmic scale factor. The
power index $\sigma $ is a (yet to be determined) function of time, which
contains information about vacuum-matter interaction. A discussion of cosmic
dynamics for $t<\tau $ requires a yet to be formulated quantum theory of
gravity and is therefore beyond the scope of this article. Note however,
that $\Lambda $ is regular at $t=0$, $\left( H\longrightarrow \infty \right) 
$, appearing to `tunnel from nothing'. Thereafter, it quickly evolves
towards its stationary point, $\frac{d\Lambda \left( t\right) }{dt}=0$.
During this ($\Lambda $-growth) period the effective equation of state $%
w\left( t\right) $ will approach $-1$ from below, to temporarily mimic a
cosmological constant. In the immediate neighborhood of $\frac{d\Lambda
\left( t\right) }{dt}=$ $\left[ \ln \left( \frac{l_{pl}}{a(t)}\right) \frac{%
d\sigma }{dt}-\sigma \left( t\right) H-\tau \frac{dH}{dt}\right] =0$, $%
\Lambda $\ is virtually constant with maximum potential energy (possibly in
the range $\langle \rho _{vac}\rangle \ =\frac{\Lambda }{8\pi G}\sim
\;10^{94}\,g\,cm^{-3}$). Such conditions will give rise to cosmic inflation,
in the early universe$.$ As the Hubble time $H^{-1}$ grows, the quantity $%
e^{-\tau H}$ quickly approaches saturation $e^{-\tau H}\rightarrow 1$,
subsequent upon which the dynamics of the universe becomes increasingly
classical, being driven by the $a(t)^{-\sigma \left( t\right) }$ part of $%
\Lambda $. In the absence of a quantum theory of gravity, it is this latter
(post-inflationary) phase that the present article addresses. In this
classical regime the evolutionary behavior of $\rho _{v}$ can be expressed
as 
\begin{equation}
\rho _{v}\left( t\right) =\frac{\Lambda \left( t\right) }{8\pi G}=\left( 
\frac{a\left( t_{0}\right) }{a\left( t\right) }\right) ^{\sigma \left(
t\right) }\rho _{v}^{\left( 0\right) },  \tag{2.2}
\end{equation}%
where $\rho _{vac}^{\left( 0\right) }$ and $a\left( t_{0}\right) $ are the
current values of dark energy density and scale factor, respectively. We are
interested in determining the functional form of the power index $\sigma
\left( t\right) $ for the evolution of $\rho _{v}$ which is consistent with
the observations of a currently accelerating universe.

\subsection{Energy equations}

In setting up the relevant energy equations we consider the dynamical
evolution of a self-gravitating cosmic medium, consisting of a two-component
perfect fluid. The total energy momentum tensor $T_{\mu \nu }$ for all the
fields is given by 
\begin{equation}
T_{\mu \nu }=\ ^{\left( m\right) }T_{\mu \nu }+\ ^{\left( vac\right) }T_{\mu
\nu }=\left[ \rho +p\right] v_{\mu }v_{\nu }+pg_{\mu \nu },  \tag{2.3}
\end{equation}%
where $\ ^{\left( m\right) }T_{\mu \nu }$ and $\ ^{\left( vac\right) }T_{\mu
\nu }$ are, respectively, the matter and the cosmic background vacuum (dark)
energy contributions to $T_{\mu \nu }$, while $\rho =\rho _{m}+\rho _{v}$,\ $%
\ p=p_{m}+p_{v}$ and $v_{\mu }$ is the 4-velocity. The total cosmic energy
is conserved, $v_{\mu }T_{\;\;;\nu }^{\mu \nu }=0$, which leads to the
standard continuity equation 
\begin{equation}
\left[ \dot{\rho}_{m}+\left( \rho _{m}+p_{m}\right) \theta \right] +\left[ 
\dot{\rho}_{v}+\left( \rho _{v}+p_{v}\right) \theta \right] =0,  \tag{2.4}
\end{equation}%
where $\theta =v_{\;;\alpha }^{\alpha }$ is the fluid expansion parameter
and $\dot{\rho}$ is the derivative taken along the fluid worldline,$\ \dot{%
\rho}=v^{\alpha }\nabla _{\alpha }\rho $. However, because of the assumed
interacting nature of dark energy in this treatment, the individual
components $^{\left( m\right) }T_{\mu \nu }$ and $^{\left( vac\right)
}T_{\mu \nu }$ are, in general, not conserved\footnote{%
The idea of an interaction between matter fields and the background was
first suggested in the early 70s by Rustal [11].}. This is a consequence of
the \textit{Cosmic Equilibrium Conjecture. }When dominant, the background
vacuum energy will act as a source of dissipative processes, while the
matter component acts as a sink of such processes. By implication, one can
write 
\begin{equation}
v_{\mu }~^{\left( vac\right) }T_{\ \ ~;\nu }^{\mu \nu }=-v_{\mu }~^{\left(
m\right) }T_{\ \ ~;\nu }^{\mu \nu }=\Psi  \tag{2.5}
\end{equation}%
where $\Psi >0$ is the particle source strength. Note, however, that Eq. 2.5
is still consistent with Eq. 2.4. More explicitly Eq. 2.5 can be split into
a source equation, 
\begin{equation}
\dot{\rho}_{v}+\left( \rho _{v}+p_{v}\right) \theta =\pi _{c}\theta , 
\tag{2.6a}
\end{equation}%
and a sink equation 
\begin{equation}
\dot{\rho}_{m}+\left( \rho _{m}+p_{m}\right) \theta =-\pi _{c}\theta , 
\tag{2.6b}
\end{equation}%
where $\pi _{c}$ is the creation pressure [3]. As Eq. 2.2 shows, solving Eq.
2.6a for the evolution of the dark energy density $\rho _{v}$ is equivalent
to solving for the parameter $\sigma =\sigma \left( a\right) $. One finds
(see [3] for details) that 
\begin{equation}
\sigma \left( \psi \right) =2+\sin 2\psi  \tag{2.7}
\end{equation}%
where $\psi $ is given by $\frac{d\psi }{da}=\left[ a\ln \left( \frac{a_{0}}{%
a}\right) \right] ^{-1}$. This leads to the solution

\begin{equation}
\rho _{v}\left( a\right) =\rho _{v}^{0}\left[ \frac{a_{0}}{a}\right]
^{\left( 2+\sin 2\psi \left( a\right) \right) }=\rho _{v}^{0}\left[ z+1%
\right] ^{\left( 2+\sin 2\psi \left( z\right) \right) },  \tag{2.8}
\end{equation}%
where the second equality is written in terms of the redshift parameter $z=%
\frac{a_{0}}{a}-1$ and now $\frac{d\psi }{dz}=\left[ \left( z+1\right) \ln
\left( z+1\right) \right] ^{-1}$. Further, one finds from these results that
the working equation of state for this interacting dark energy is 
\begin{equation}
p_{v}=-\frac{1}{3}\left( 1-\sin 2\psi \right) \rho _{vac}.  \tag{2.9}
\end{equation}%
The results in Eqs. 7 to 9 constitute, in our treatment, the formal solution
for the post-inflationary evolution of the interacting background dark
energy with the scale factor $a\left( t\right) $. The sinusoidal power index 
$\sigma \left( \psi \right) $ depicted in Eq. 2.7 accounts for the
interactions between dark energy and matter fields. These interactions
constrain the evolution of the two fields relative to each other (see Eq.
2.10). In turn, it is this feature that protects the universe from run-away
vacuum-induced accelerations. The limiting forms of the above solution (both
during the vacuum dominated and matter dominated regimes) can be used to put
bounds on the post-inflationary $\left( e^{-\tau H}\longrightarrow 1\right) $
evolution of the dark energy $\rho _{v}$ with respect to the matter fields $%
\rho _{m}$. One finds [3] that $\rho _{v}$ evolves so that 
\begin{equation}
\left( \frac{3K}{6K+2}\right) \left( 3\gamma -2\right) \rho _{m}\leq \rho
_{v}\leq \left( \frac{3K}{6K-2}\right) \left( 3\gamma -2\right) \rho _{m}, 
\tag{2.10}
\end{equation}%
where $\gamma =\left[ \frac{4}{3},1\right] $ and currently the dissipation
parameter $K$ is constrained to $\frac{1}{3}<K\lesssim \frac{4}{9}$. Through
this constrained relative evolution of the fields, (Eq. 2.10, also see [3]),
the approach explains the Coincidence Problem and leads to a predictable
future dynamics of the universe, hence offering a resolution to the \textit{%
dynamical problem. }Moreover, as the vacuum energy and matter fields
oscillate into each other (see Eq. 2.10 and 8 and also [3]), they drive the
universe into decaying oscillations about its time-evolving dynamical
equilibrium state. Thus, as the universe ages, it also moves closer to being
permanently critical.\ In this respect, it is therefore not surprising that
the observed universe looks virtually critical.

In the initial limit $K\longrightarrow 0$, there is no dissipation. Then,
and only temporarily, $\rho _{v_{\left( K=0\right) }}$ is the energy in the
form of a pure cosmological constant, and the universe must inflate.
Immediately after this however, $K$ will assume non-trivial, positive,
values and this signals the decay of $\Lambda $. Such a scenario is a
consequence of the time-asymmetry of the de Sitter state. In this respect,
the approach independently suggests inflation as a natural initial condition
for the current cosmic dynamics and, as a by-product, suggests a potential
role matter plays in the universe. The above results suggest that the
current cosmic dynamics is based on the need for the universe to seek
equilibrium conditions against run-away vacuum-induced accelerations. This,
indeed, is the content of the \textit{Cosmic Equilibrium Conjecture (CEC)}.

\section{Cosmic self-regulation}

As was pointed out earlier in this article however, CEC as applied to cosmic
dynamics so far, constitutes a one-sided part of what should be a more
general and symmetrical statement. This is because spacetime may not only be
(vacuum) accelerated but could also be (gravitationally) collapsed. To this
end, one expects that there should exist a general principle which protects
the universe from run-away behavior in both directions. In the remaining
part of this article we set up a framework for such a principle. The
framework is based on the following more encompassing statement, which (here
and henceforth) is referred to as the \textit{General Cosmic Equilibrium
Conjecture (G-CEC):}

\begin{conjecture}
The universe will resist all agents tending to move it away from dynamical
equilibrium (be they vacuum energy based or gravitationally based).
\end{conjecture}

\subsection{\protect\bigskip Case against run-away spacetime collapse}

In the rest of this article we shift our discussion to the other side of
cosmic equilibrium and search for the universe's stability to \textit{total}
or run-away gravitational collapse. Since in the physics literature there is
no known observational evidence for a previous and/or future global cosmic
collapse, we will (with no loss of generality) illustrate the validity of
G-CEC based only on local spacetime (gravitational) collapse, namely the
formation of black holes. As a useful by-product such an approach also will
have the advantage of directly linking global cosmic dynamics to local
spacetime (black hole) dynamics. In hind sight, such a link is natural in
the sense that spacetime here is no longer simply `a playing field' on which
particle dynamics is described but is, instead, what is being either
stretched by the vacuum energy or collapsed by gravity. In other words
spacetime becomes a dynamical variable\textit{. }

As we move to apply these concepts to gravitational collapse, we should
briefly mention one issue that would seem to make such an extension
difficult and later it will be pointed out how nature overcomes the
difficulty. In [3] the original CEC was based on the argument that during
periods of cosmic acceleration resulting from vacuum domination, increase in
cosmic internal energy (i.e. matter creation) must be a one-way process.
There, vacuum energy could create matter but not vise versa. This is because
(under the conditions) a reverse process would globally violate the entropy
law $s_{\;;\mu }^{\mu }.\geq 0$ and hence tend to rotate the thermodynamic
arrow of time. The one-way creation feature is reflected in the solution
(Eqs. 7 to 8) obtained previously for the evolution of the interacting
vacuum energy density $\rho _{v}$. In this solution, the oscillations appear
only in the power index $\sigma \left( a\right) $; and since manifestly $%
\sigma \left( \psi \right) =2+\sin 2\psi \left( a\right) >0,~\forall a\left(
t\right) $, then the evolutionary slope of $\rho _{v}$ always takes one, 
\textit{and only one}, sign i.e. negative. As a result the decaying vacuum
energy is always time-wise single valued. The alternative, in which vacuum
energy $\rho _{v}$ would be multi-valued (i.e. `creatable' from matter
fields) would violate the entropy law. As we generalize CEC to G-CEC to
include effects of gravitational collapse, we will revisit this question to
find that with regard to gravitational collapse, nature renders moot the
issues pertaining to possible violations of \ both the entropy law and the
thermodynamic arrow of time.

The concept of gravitational collapse has effectively been with us since
1916 when Schwarzschild first presented a solution, to the Einstein field
equations, for the gravitational field of a spherically symmetric mass. To
date, the physics of these end-products has evolved both in breadth and
depth, both theoretically and observationally, so much so that black holes
are currently considered virtually discovered. The boundary of a black hole,
a 2-sphere (geometrical singularity in Schwarzschild coordinates), provides
(at least currently) the observational limit with regard to the dynamical
evolution of the hole. It is, however, widely believed that during
gravitational collapse of a spacetime region to form a black hole, the
matter itself continues to collapse beyond the horizon towards a physical
singularity. This perception has been motivated, from a theoretical view
point,\ by the apparent absence of any known force that would otherwise
counteract the action of gravity under such circumstances. Inquiries in the
nature of spacetime singularities led to the famous Singularity Theorems
[12]. It was conjectured then, as a Cosmic Censorship [13], that nature
shielded the would-be naked singularities from exposure to external
observers. To date, it is still widely believed that in spite of limitations
relating to their observability, spacetime singularities should exist (naked
or not)\footnote{\textit{The debate (and bet) that was later to develop
between Stephen Hawking and Kip Thorne, on the necessity (or not) of cosmic
censorship is known even well beyond the relativistic astrophysics community.%
}}.

The classical singularity theorems [12] are built on the assumption that the
physical fields involved obey the dominant energy condition, DEC, i.e. $%
\varrho \geq 0$ and $\rho +3p\geq 0$ (for a review of these conditions, see
e.g. [14]). On the other hand, gravitationally collapsing matter does
exhibit two characteristics which become important in the proceeding
discussion, in so far as they create conditions that eventually violate the
dominant energy condition. These characteristics relate to (i) loss of
particle kinetic energy from the collapsing system; (ii) losses in internal
degrees of freedom in the collapsing system. To illustrate the importance of
these points we begin with a familiar example, namely the formation of a
neutron star, because here the initial signs of the above two effects can be
observationally inferred. During the formation of a neutron star, it is
believed that most of the orbital electrons do smash into the protons of the
host nuclei through the inverse $\beta $-decay process $\left(
e^{-}+p\longrightarrow n+\nu _{e}\right) $. This process will increase the
neutron content of the collapsing star while at the same time carrying away
most of the particle kinetic energy and internal degrees of freedom (e.g.
spin) through the escaping neutrinos $\nu _{e}$. For larger stars collapsing
into black holes, the losses of such degrees of freedom is expected to
become much more extreme, leading to a growth of degeneracy in the particle
states of the collapsing system. Moreover, the loss of particle kinetic
energy will compound the situation by evolving such a `degenerating' system
towards a Bose-Einstein-like condensate. In case of a black hole such losses
could be associated with Hawking radiation. In retrospect, this behavior is
consistent with the "no hair theorem" [15]. The classic black hole manifests
its presence only through its bulk (degenerate) mass and bulk angular
momentum (individual particles degrees of freedom in the system having been
compromised). Eventually, as a significant number of individual particles
inside the black hole lose self-identity, the two effects will render the
bulk matter (at the collapse core) to undergo a phase transition, taking on
the form of a scalar field. The fraction of the original collapsing matter
that will be in this scalar form will depend, among other things, on the
stage of collapse. In any case, a rigorous field-theoretic description of a
such scenario will certainly require a quantum theory of gravity, which is
not in place yet. In this connection, however, one can already sense that
the vacuum-matter interactions implied here and in the earlier discussion
will likely require a theory involving higher order terms of the curvature
scalar $R$ and for which the Einstein theory, $R_{\mu \nu }-\frac{1}{2}%
Rg_{\mu \nu }=T_{\mu \nu }$, is a low energy limit\footnote{%
The Einstein theory of gravity does not discuss vacuum-matter interactions
and/or oscillations since there $\Lambda $ is cosnstant.}.

Our present interest is not to derive such a theory; rather we only seek to
validate G-CEC by demonstrating that it leads to results consistent with
observations, helps explain existing paradoxes and predicts, among other
things, a self-regulating cosmic dynamics. It will therefore suffice to keep
our discussion simple but illustrative.\textit{\ }To this end, one finds
that for illustrative purposes, one can still bring out those effects of
gravitational collapse, consistent with the preceding argument, by simply
treating the inner core of the collapsed system as a minimally coupled
scalar field $\phi $, with the familiar action [16],

\begin{equation}
S=\int d^{4}x\sqrt{-g}\left[ \frac{1}{2}g^{\mu \nu }\partial _{\mu }\phi
\partial _{\nu }\phi -V\left( \phi \right) \right] ,  \tag{3.1}
\end{equation}%
where $g_{\mu \nu }$, in this case, is the local metric tensor of the
spacetime inside the black hole, in the relevant region occupied by $\phi $.
The associated energy-momentum tensor is

\begin{equation}
T_{\mu \nu }=\frac{1}{2}\partial _{\mu }\phi \partial _{\nu }\phi +\frac{1}{2%
}g^{\alpha \beta }\partial _{\alpha }\phi \partial _{\beta }\phi -V\left(
\phi \right) g_{\mu \nu }.  \tag{3.2}
\end{equation}%
As pointed out earlier, loss of virtually all the particle kinetic energy
through escaping relativistic particle fluxes leaves behind a one-state
frozen condensate. This implies for the system that $\partial _{\mu }\phi =0$%
, which effectively leaves the energy-momentum tensor of the scalar field
part (Eq. 3.2) to be only proportional to the local spacetime metric $g_{\mu
\nu }$, and one can write%
\begin{equation}
T_{\mu \nu }=-\bar{\rho}_{v}g_{\mu \nu }.  \tag{3.3}
\end{equation}%
Clearly, Eq. 3.3 represents the energy momentum tensor corresponding to a de
Sitter geometry with a positive energy density $\bar{\rho}_{v}$, and
negative pressure components $\bar{p}_{v}=-\bar{\rho}_{v}$. The latter
provide the spacetime in consideration with an anti-gravitating character.\
During gravitational collapse, the amount of regular matter turning into
this de Sitter-like fluid will grow, and do so at the expense of the regular
matter available. The resulting increase in the negative pressure will slow
down the overall gravitational collapse, eventually compromising it all
together when $T_{\mu }^{\mu }\left( vac\right) \sim T_{\mu }^{\mu }\left(
matter\right) $ in the region. This suggests, contrary to the traditional
view, that gravitational collapse will not continue all the way to form a
physical singularity. Consequently, according to the \textit{General Cosmic
Equilibrium Conjecture}, black holes do not form spacetime singularitities.
And since black holes are the only known candidates that could induce such
extreme spacetime curvature, it appears that nature does not just abhor
`naked' singularities but rather (nature) abhors all physical singularities
altogether, naked or not.

One notes a symmetry that has so far developed in this discussion. Clearly,
the process of matter turning into a cosmological constant-like vacuum
energy, to offset the formation of spacetime singularities, is exactly
opposite to the process discussed earlier in the article, in which a
dominating vacuum energy will decay into (matter) fields that increase
cosmic inertia and compromise the cosmic acceleration. Such a symmetry
should, therefore, rotate vacuum energy into matter fields and vise versa,
and should admit a duality between positive and negative spacetime
curvature. The symmetry implied in the vacuum-to-matter and matter-to-vacuum
processes suggests some deeper principle underlying the dynamics of the
universe. A rigorous discussion of such a principle and the group-theoretic
characteristics of the implied symmetry behind it is not treated in the
current article and will be carried on elsewhere. Nevertheless, the
inference here is that the universe is stable to, and hence self-regulating
against, irreversible conditions arising from either extreme vacuum or
positive curvature domination. This is the content of the \textit{General
Cosmic Equilibrium Conjecture}.

One may at this point wonder how it is that in turning matter fields into a
de Sitter vacuum in black holes, nature would not violate the entropy law.
The answer is two fold. First, one recalls that the late time process when
such a transition begins to occur is already hidden from the external
observer by the black hole horizon so that any such violations are not
physically observable! In this regard, black hole horizons become inevitable
natural participants in the "regulation conspiracy" process of spacetime
dynamics. Moreover, any such horizon is endowed with a temperature $T_{H}$
and an associated radiation flux directed towards the external observer.
Such radiation flux carries entropy with it. The result is that from the
vintage point of the external observer, the process does instead result in
the increase of entropy. Consequently, the entropy law is still protected
here and, manifestly, the natural existence of the black hole horizon
renders the issue moot. The universe may indeed be `wiser' than it is often
given credit for!

The notion that a black hole may not contain a spacetime singularity is not
new. What is new here is putting forward a physically motivated
justification as to why black holes should not contain such singularities
and further tying this up (through G-CEC) to a bigger cosmic picture of a
universe that self-regulates its dynamics. Since they were first suggested
by Gliner [17], non-singular black holes, along with other related
end-products, have been discussed for a while now, by several authors
including [18], [19], [20], [21], [22]. In [18], Brandenberger first
constructed an effective action based on higher derivative modification of
Einstein's theory, in which all homogeneous solutions are nonsingular and
asymptotically approached the de Sitter space for $r\longrightarrow 0$. This
work was later extended [20] to study scenarios in which a universe can be
born from a black hole resting in a parent universe. More recently, Hawking
radiation of such a nonsingular black hole has been studied [22] by Easson.
Moreover, in her works, Dymnikova has consistently (see [19] and citations),
argued that the final end-product of gravitational collapse should be a
nonsingular black hole with an $r$-dependent $\Lambda \left( r\right) $ de
Sitter-like core and with an energy-momntum tensor falling in the Petrov
classification [(II)(II)]. On the other hand, in [21] Mazur and Mottola
discuss gravitational collapse in which the end-product with a de Sitter
core does not form a horizon. All these works clearly indicate that
nonsingular gravitational collapse is becoming of interest to relativistic
astrophysics and cosmology. One hopes that the physical justification for
nonsingular collapse suggested in the present article will further stimulate
the debate on the issue.

\section{Conclusion}

In conclusion we have, in this article, proposed that the universe is
self-regulating in its dynamics. The framework implies that the universe
dictates its exit to inflation, on the one extreme, while on the other
extreme it negates gravitational curvature domination and the formation of
local spacetime singularities. The central feature of the proposal, which is
expressed in the \textit{General Cosmic Equilibrium Conjecture (G-CEC),} is
that the universe always acts to off-set any tendencies to shift it away
from dynamical equilibrium. It achieves this through a mechanism of
interaction between the background vacuum energy and matter fields. We have
shown, on the one hand, that whenever the universe is vacuum dominated, the
consequent spacetime acceleration leads to creation of physical fields that
increase its internal energy and hence its gravitational content. On the
other hand, whenever spacetime is extremely matter dominated and faced with
the prospect of total collapse the matter at the core of the resulting black
hole undergoes a phase transition to a de Sitter-like vacuum, hence avoiding
formation of a physical singularity. The model relates global cosmic
dynamics with local spacetime (black hole) dynamics suggesting a deeper
symmetry underlying cosmic dynamics. In essence this relation is natural
when one considers that spacetime (in both cases) is no longer just a
"playing field" for particle dynamics but is, indeed, itself dynamical. The
approach, therefore, suggests some (yet to be explored) principle and its
symmetry that governs cosmic dynamics, directing the universe away from
extreme irreversible conditions, by facilitating the interaction of the
fields involved. The scenario reflects a universe which is always in search
of equilibrium. This is the content of G-CEC. To our knowledge, this is the
first time in the literature that the existence of a common principle behind
both global cosmic dynamics and local spacetime dynamics, and behind both
spacetime acceleration and spacetime collapse, has been suggested.

As we summarize below, the model (A) re-traces the evolution of the universe
in a way that is consistent with existing observations; (B) provides a
natural explanation to existing problems/puzzles; (C) makes predictions.

\begin{quotation}
A) \textbf{Consistency: }\textit{the model}

(1) reiterates the early universe was dominated by a vacuum energy that set
it in an inflationary acceleration;

(2) suggests that spacetime backreacts to this acceleration by creating
radiation/matter fields that increase its gravitational content;

(3) holds that (i) A2$\Longrightarrow $ justification for cosmic exit out of
inflation; (ii) A2$\Longrightarrow $ increased cosmic radiation after
inflation.

B)\textbf{\ Explained Puzzles: }\textit{the model implies that}

(1) since a de Sitter spacetime is time-asymmetric and radiative, or (as a
corollary) since vacuum energy acting as cosmic fuel must degrade into
matter fields, then the question of why $\Lambda $ is small now is explained
by its inevitable decay\footnote{%
Note that this process is energy conserving and at any time all the initial
energy in a cosmological cosnstant is accounted for.}.

(2) since in a decaying-vacuum dominated universe matter creation is a
one-way process, it is seen why the amplitudes of the densities $\rho _{m}$
and $\rho _{v}$ do approach each other as the universe expands and ages,
which explains the Coincidence Problem;

(3) the premise that the universe is equilibrium seeking coupled with the
observation that the amplitudes of the field densities approach each other
(see (4) explains why the universe on the average appears critical and will
appear more so in future.

\ C) \textbf{Predictions:} \textit{the model predicts that}

(1) the assumption of an early inflationary phase, coupled with decay of $%
\Lambda $ in an expanding universe is necessarily followed by a matter
domination phase (see A1,2,3) which, in turn, necessarily leads to the
current low density vacuum domination and cosmic acceleration. Conversely,
the observed current low level acceleration predicts a previous matter
dominated phase, preceded by an early vacuum dominated inflationary phase;

(2) the current and any other (prior/consequent) cosmic accelerations must
be variable (time-dependent) due to the vacuum-matter oscillations;

(3) consistent with C1 and C2, there is a future exit to the current
acceleration through creation of long wavelength modes as a backreaction;

(4) there will be future periods of alternating vacuum and matter domination
with decaying and time-wise more comparable maxima in the field densities,
which offers an explanation to the \textit{dynamical problem};

(5) the universe does not admit creation of \ (physical) spacetime
singularities and, in particular, black holes do not contain such
singularities.
\end{quotation}

\begin{acknowledgement}
I would like to thank Robert Brandenberger, Ron Mallet and Fred Adams for
various discussions and comments. This work was made possible by funds from
RIT
\end{acknowledgement}

\end{document}